\documentclass[numbers]{elsarticle}

\usepackage[top=1.25in, bottom=1.25in, left=1.5in, right=1.5in]{geometry}
\usepackage{lineno}
\usepackage{soul}
\usepackage{color}
\usepackage{tikz}
\usetikzlibrary{quotes,angles}
\usepackage[pagebackref=true,
            colorlinks=true,
            bookmarks=true,
           ]{hyperref}   
           
\usepackage[most]{tcolorbox}           
\usepackage{graphicx}
\usepackage{epstopdf}
\DeclareGraphicsExtensions{.eps}
\usepackage{siunitx}

\usepackage{empheq}
\usepackage{amssymb}
\usepackage{array}
\usepackage{amsmath, scalerel}
\usepackage{leftidx}
\usepackage{amssymb}

\usepackage{amsthm}

\usepackage{mathptmx}

\usepackage{amsfonts}
\usepackage{multicol}
\usepackage{mathrsfs}
\usepackage{tensor}
\usepackage{subfig}
\usepackage{hhline}
\usepackage{upgreek}
\usepackage{cancel}
\usepackage{ulem}

\usepackage{multirow}
\usepackage{setspace}

\usepackage[flushleft]{threeparttable}
\usepackage{makecell,booktabs}

\usepackage{amsmath,scalerel}

\newcommand{\tightoverset}[2]{\mathop{#2}\limits^{\vbox to -.5ex{\kern-0.75ex\hbox{$#1$}\vss}}}
\modulolinenumbers[5]

\journal{Elsevier}

\begin{document}

\allowdisplaybreaks[4]

\begin{frontmatter}
\title{An effective anisotropic visco-plastic model dedicated to high contrast ductile laminated microstructures: Application to lath martensite substructure }

\author[1,2]{V. Rezazadeh}
\author[1]{\corref{cor}R.H.J. Peerlings}\ead{r.h.j.peerlings@tue.nl}
\author[3]{F. Maresca}
\author[1]{J.P.M. Hoefnagels}
\author[1]{M.G.D. Geers}

\address[1]{Department of Mechanical Engineering, Eindhoven University of Technology (TU/e), P.O.Box 513, 5600 MB Eindhoven, The Netherlands}
\address[2]{Materials Innovation Institute (M2i), P.O.Box 5008, 2600 GA Delft, The Netherlands}
\address[3]{Faculty of Science and Engineering, P.O.Box 72, 9747 AG  Groningen, The Netherlands}

\cortext[cor]{Corresponding author.}

%% Abstract
    \begin{abstract}
	In particular types of layer- or lamellar-like microstructures such as pearlite and lath martensite, plastic slip occurs favorably in directions parallel to inter-lamellar boundaries. This may be due to the interplay between morphology and crystallographic orientation or, more generally, due to constraints imposed on the plastic slip due to the lamellar microstructural geometry. This paper proposes a micromechanics based, computationally efficient, scale independent model for particular type of lamellar microstructures containing softer lamellae, which are sufficiently thin to be considered as discrete slip planes embedded in a matrix representing the harder lamellae. Accordingly, the model is constructed as an isotropic visco-plastic model which is enriched with an additional orientation-dependent planar plastic deformation mechanism. This additional mode is activated when the applied load, projected on the direction of the soft films, induces a significant amount of shear stress. Otherwise, the plastic deformation is governed solely by the isotropic part of the model. The response of the proposed model is assessed via a comparison to direct numerical simulations (DNS) of an infinite periodic two-phase laminate. It is shown that the yielding behavior of the model follows the same behavior as the reference model. It is observed that the proposed model is highly anisotropic, and the degree of anisotropy depends on the contrast between the slip resistance (or yield stress) of the planar mode versus that of the isotropic part. The formulation is then applied to model the substructure of lath martensite with inter-layer thin austenite films. It is exploited in a mesoscale simulation of a dual-phase (DP) steel microstructure. The results are compared with those of a standard isotropic model and a full crystal plasticity model.  It is observed that the model can reproduce the characteristic material behavior of the latter while keeping the computational cost comparable to the former.
    \end{abstract}

\end{frontmatter}

%\linenumbers

%========================================================================================================================================================================
\section{Introduction}

	Lamellar structures are known to be an essential building block of various materials. On different scales, they are either formed to minimize the total elastic energy, e.g., in systems with structural variants such as, pearlite \citep{queiros2019metallographic, VERMEIJ2022114327}, intermetallic \textit{Ti-Al} alloys \citep{Yamaguchi2000,Huang2018}, nano-bainitic structures \citep{Bhadeshia2013, Caballero2019}, and lath martensite \citep{Morito2003}, or manufactured to improve mechanical properties, e.g., in thermally-sprayed coating materials \citep{Santana2008, KLUSEMANN20121828}, and semi-crystalline polymers \citep{Yoon1979, Bartczak1992}. Highly anisotropic behavior, originating from the different types of crystallographic or morphological heterogeneities of the phases within the laminate, is a typical characteristic of these structures. It has been observed that in \textit{Ti-Al} alloys the lamellar interfaces can act as obstacles to the homogeneous plastic deformation, resulting in channeling of plasticity and hence, an easy deformation mode parallel to the interfaces \citep{Paidar2007}. The molecular chain axes (chain direction) are constrained slip directions in semi-crystalline polymers, forcing extra kinematical constraints on the crystalline phase \citep{Bartczak1992}. 
	
	A prime example of inhomogeneous plastic activity, and the main motivation for the model developed in this paper, is the lamellar microstructure of lath martensite \citep{DU2018411}. There is increasing experimental evidence that large plastic strains occur in the lath martensite \citep{DU2016117, KWAK2016104}, even though it is conventionally be regarded as a brittle phase. It has been hypothesized that this plastic deformation, which occurs more favorably parallel to the lath boundaries \citep{DU2019107646}, is facilitated by thin retained austenite films that trapped between the laths \citep{Maresca2014}. These films, by virtue of the orientation relationships between the martensite laths and their parent austenite, have slip systems that are parallel to the lath, and which hence glide easily. Alternatively, the remarkable ductility in particular directions has been attributed to the morphology of the laths which typically have an aspect ratio of $30:1$, in the length direction (habit plane direction) compared to the thickness direction, and hence the dislocations on the slip systems parallel to the habit plane can glide comparatively freely \citep{MORSDORF2016202, RYOU2020139090}. Regardless of the exact physical mechanism underlying this phenomenon, plastic activity/slip is mediated favorably in directions parallel to the lath boundaries \citep{DU2019107646, INOUE2019129}, which results in a significant degree of anisotropy in the plastic response. 
    
    Detailed, full-scale models of the lamellar microstructure of small volumes (a few grains) may give much insight on the relevant mechanisms discussed above \citep{HatemAndZikry2009, GHAFFARIAN2022153439}. However, mesoscale models, i.e. models containing many grains, cannot resolve the length scale of the inter-granular lamellar microstructures in full detail, as such models would entail a prohibitive computational cost. 
    Therefore, one typically resorts to homogenization to obtain (or compute online) the response of the laminate.
       
	 From highly-idealized analytical and semi-analytical models of \citet{reuss1929berechnung}, \citet{voigt1908lehrbuch}, \citet{hashin1961note} to variational multiscale models such as first and second-order homogenization approaches \citep{KOUZNETSOVA20045525, Geers2010}, numerous homogenization techniques have been developed to predict the effective response of multiphase structures. The attention of this work, however, is restricted to z particular class of methods addressing lamellar structures. Homogenization of laminates generally involves incorporating the behavior of two (or more) phases with the corresponding interaction(s) between them \citep{livingston1957multiple, kocks1981many}. Neglecting the interactions can result in too stiff (Taylor-type models \citep{taylor1938plastic, hutchinson1976bounds}), or too soft (Sachs-type models \citep{Sachs}) behavior. To satisfy compatibility and equilibrium across the interfaces of phases/grains, intermediate models such as relaxed constrained type models or self-consistent models have been developed \citep{lebensohn1993self, MOLINARI19872983, molinari1997self}. Similar concepts have also been used to model the interaction between pairs of grains in non-lamellar microstructures such as polycrystals, e.g. LAMEL and ALAMEL models \citep{van2005deformation}. In these models, the representative volume is assumed to be a single crystal/grain/inclusion in a homogeneous equivalent medium. \citet{lee1993micromechanical} proposed a visco-plastic two-phase composite inclusion model for deformation and texture evolution in semi-crystalline polymers (with a lamellar structure of the crystalline and amorphous phases). Later they adopted the idea of using a two-phase composite inclusion for FCC polycrystals, where the composite inclusion is represented by a bicrystal with arbitrary misorientation and with assumed extended planar interface \citep{Lee2002}. \citet{van2003micromechanical} extended this model to an elasto-viscoplastic formulation. The idea of two-phase laminates is used in \citet{ORTIZ20002077} for describing the lamellar dislocation structures which develop at large strains. Microstructures of lamellar type have long been treated within the context of the crystallographic theory of martensitic transformation \citep{pedregal1993laminates, bhattacharya1991wedge, STUPKIEWICZ2006126}. Based on the concept of a laminate composed of a martensite plate and an austenite layer, \citet{KOUZNETSOVA2008641} modeled the transformation plasticity of variant formation in the martensitic transformation. \citet{maresca2016reduced} proposed a reduced crystal plasticity model to account for the limited number of active slip systems in the austenite layers trapped between the laths of martensite, and validated the model using lath martensite microstructures. Finally, \citet{KLUSEMANN20121828} proposed a homogenization method in a small strain setting to capture the material behavior of two-phase laminates characterized by a thin-layer-type microstructure found in thermally-sprayed coating materials like \textit{WC/Fe}.
  	     
   Here, we focus on the plasticity of two-phase laminate-type microstructures, and particularly on the cases where one of the phases is sufficiently thin such that it can be considered as a discrete plane, cf. the boundary sliding mechanism in lath martensite discussed above. Taking benefit of this feature, we model the system as a comparatively hard, isotropic elasto-viscoplastic matrix in which a family of parallel soft discrete plastic planes are embedded. The planes may exhibit viscoplastic sliding which is assumed to be isotropic along the plane and which is driven by the resolved shear stress vector. Within their planes, these films experience the same stress as the matrix. However, they respond to it differently, depending on their properties and their orientation with respect to the loading direction. This additional visco-plastic deformation mode is added directly to the plastic deformation rate of the matrix, resulting in an effective model which has a much lower complexity and computational cost than the homogenized approaches discussed above, e.g., that by \citet{maresca2016reduced} for lath martensite.
   
	In what follows we first, in Section \ref{sec:ModelFormulation}, formulate the model as sketched above in a three dimensional finite deformation setting. In Section \ref{sec:ModelBehavior}, the response of the model is assessed via a comparison to direct numerical simulations (DNS) carried out on an infinite periodic two-phase laminate. Furthermore, in this section, the influence of the film spacing is investigated. Section \ref{sec:application} presents the bicrystal laminate of martensite laths and inter-lath retained austenite, which reflects the fully resolved model of lath martensite. This model is used as a reference when applying our new, effective laminate model to the case of lath martensite. Then, the model is implemented in a FFT based spectral solver to be used in mesoscale simulations. An RVE of a dual-phase (DP) microstructure with $40\%$ of martensite is simulated, and the results are compared with those of a standard isotropic and a crystal plasticity model. In the end, the computational gain of the model is  discussed shortly by comparing to the model of lath martensite proposed by \citet{Maresca2014}.  

	Throughout the paper, the notations $a$, $\mathbf{b}$, $\mathbf{C}$, and $\mathbb{D}$ denote scalars, vectors, second-order tensors and fourth-order tensors, respectively. Single and double contractions are denoted by "$\cdot$" and "$:$", respectively. The inner product between two second-order tensors is given by, $\mathbf{C}=\mathbf{A}\cdot\mathbf{B}$ as $C_{ik}=A_{ij}B_{jk}$ and the double inner product of a fourth-order tensor with a second-order tensor is defined by $\mathbf{C}= \mathbb{A}:\mathbf{B}$ implies $C_{ij}= A_{ijkl}B_{lk}$. The action $|| \cdot||_{F}$ designates the \textsc{Frobenious} norm, i.e. $|| \mathbf{A} ||_{F}=\sqrt{\mathrm{tr}(\mathbf{A}\cdot\mathbf{A}^T)}$, where '$\mathrm{tr}$' is the trace operator, and symbol '$T$' indicates transposition. The tensor (or dyadic) product between two vectors is denoted by $\mathbf{C}=\mathbf{a} \otimes \mathbf{b}$ ($C_{ij}=a_{i}b_{j}$).

%
%========================================================================================================================================================================
\section{Model formulation}
\label{sec:ModelFormulation}
\begin{figure}[ht!]
\centering
  \includegraphics[width=0.9\linewidth]{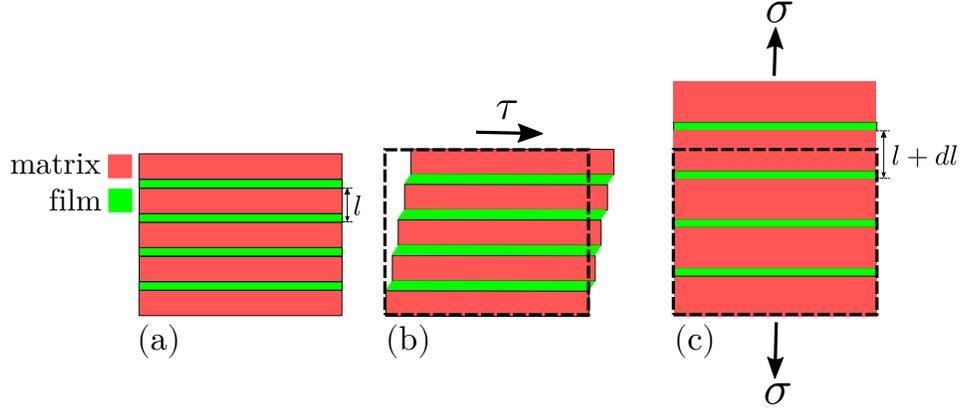}
  \caption{a) Sketch of a laminate consisting of the matrix phase with embedded soft thin films. b) Simple shear applied parallel to the plane resulting in the accommodation of all slip by the films only. c) Tension applied perpendicular to the films deforming only the matrix phase. }
  \label{fig:laminate}
\end{figure}

	Figure \ref{fig:laminate}a shows a sketch of an infinite periodic two-phase laminate obtained by embedding infinitesimally thin soft films in a harder, continuous matrix. The elastic response of the model is governed by the homogeneous matrix phase, whereas the plasticity is composed of two separate contributions: an isotropic visco-plastic part representing the matrix behavior, enhanced by a planar isotropic sliding mechanism that models the embedded films. The planar model captures a specific orientation-dependent plastic deformation mode that the composite may exhibit due to the presence of the softer thin films. In the case of shear exerted parallel to these planes, see Figure \ref{fig:laminate}b, they are expected to accommodate virtually all of the plastic strain, while applying stress in any other direction the harder matrix involves, resulting in a harder overall response. In the particular case of tension applied perpendicular to the films, Figure \ref{fig:laminate}c, the films are inactive and the response is governed solely by the standard isotropic (matrix) part of the model.
%%-----------------------------------------------------
%%-----------------------------------------------------

%------------------------------------------------------------------------------------------------------------------------------------------------------------------------
\subsection{Elasticity}
	The deformation of the laminate is characterized by the deformation gradient tensor, $\mathbf{F}$ which can be split multiplicatively into its elastic contribution, $\mathbf{F}_{\mathrm{e}}$, and a plastic contribution $\mathbf{F}_{\mathrm{p}}$, as follows,
%%-----------------------------------------------------
\begin{align}\label{eq:1}
\mathbf{F}= \mathbf{F}_{\mathrm{e}} \cdot \mathbf{F}_{\mathrm{p}}.
\end{align}
%%--------------------------------------------
%%--------------------------------------------
	$\mathbf{F}_{\mathrm{p}}$ represents an isochoric lattice preserving plastic deformation. The elastic deformation and rigid-body rotation are encoded in $\mathbf{F}_{\mathrm{e}}$. The stress induced by the elastic deformation is expressed via generalized \textsc{Hooke}'s law as:
%%-----------------------------------------------------
\begin{align}
\mathbf{S}_\mathrm{e}= \mathbb{C}:\mathbf{E}_\mathrm{e},
\end{align}
%%-----------------------------------------------------
	where $\mathbb{C}$ denotes the 4th-order isotropic elasticity tensor, and $\mathbf{E}_e=1/2(\mathbf{C}_\mathrm{e}-\mathbf{I})$ is the elastic \textsc{Green-Lagrange} strain tensor, in which $\mathbf{C}_\mathrm{e}= \mathbf{F}_{\mathrm{e}}^{T}\cdot\mathbf{F}_{\mathrm{e}}$, is the elastic right \textsc{Cauchy-Green} deformation tensor. $\mathbf{S}_\mathrm{e}$ is the push forward of the second \textsc{Piola-Kirchhoff} stress $\mathbf{S}$ to the intermediate (plastic) configuration, i.e. $\mathbf{S}_\mathrm{e}=\mathbf{F}_\mathrm{p}\cdot \mathbf{S} \cdot \mathbf{F}_\mathrm{p}^{T}$. The elastic response of the model is entirely governed by the matrix, and the films contribute only to the plastic part of the deformation. Accordingly, $\mathbb{C}$ represents the elastic stiffness of the matrix.

%------------------------------------------------------------------------------------------------------------------------------------------------------------------------
\subsection{Plasticity}
	The model incorporates extra discrete planar modes into the plastic response of the matrix. As a result, the plastic velocity gradient in the model is additively split into two separate contributions,
%%-----------------------------------------------------
\begin{align}\label{VelocityGradient}
\mathbf{L}_{\mathrm{p}} = \dot{\mathbf{F}}_{\mathrm{p}}\cdot \mathbf{F}_{\mathrm{p}}^{-1} = \mathbf{L}_{\mathrm{p}}^{\mathrm{m}} +  \mathbf{L}_{\mathrm{p}}^{\mathrm{f}},
\end{align}
%%-----------------------------------------------------
	in which $\mathbf{L}_{\mathrm{p}}^\mathrm{m}$ represents the plasticity governed by the matrix and $\mathbf{L}_{\mathrm{p}}^{\mathrm{f}}$ accounts for the effective sliding of the films. 	

%%----------------------------------------------------------------------------------------------------------------------------------------------------------------------
\subsubsection{Visco-plasticity of the matrix}\label{sub:matrixplasticity}
	First, we review the standard isotropic visco-plasticity formulation with isotropic hardening model proposed by \citep{ANAND1985213}. The associated plastic velocity gradient for the matrix phase, $\mathbf{L}_{\mathrm{P}}^\mathrm{m}$, is given by,
%%-----------------------------------------------------
\begin{align}\label{Lp3D}
\mathbf{L}_{\mathrm{P}}^\mathrm{m}= \dot{\gamma}^{\mathrm{m}}\dfrac{\mathbf{M}^{\mathrm{dev}}}{|| \mathbf{M}^{\mathrm{dev}} ||_{F}}. 
\end{align}
%%-----------------------------------------------------	
	Where $\dot{\gamma}^{\mathrm{m}}$ is based on the widely adopted power-law relation for the plastic shearing rate \citep{hutchinson1976bounds} in terms of the second invariant of the deviatoric \textsc{Mandel} stress tensor, $\mathbf{M}= \mathbf{S}_\mathrm{e} \cdot \mathbf{C}_\mathrm{e}$, as: 
%%-----------------------------------------------------
\begin{align}\label{ShearMatrix}
\dot{\gamma}^{\mathrm{m}}=\dot{\gamma}_0\left( \sqrt{\dfrac{1}{2}}  \dfrac{ || \mathbf{M}^{\mathrm{dev}} ||_{F}
}{\tau_{\mathrm{y}}^{\mathrm{m}}} \right)^{1/n},
\end{align}
%%-----------------------------------------------------
 where $\dot{\gamma}_0$ is reference plastic strain rate, and $n$ is strain rate sensitivity parameter. Note that here we use the shear equivalent \textsc{von Mises} form, for consistency with the planar plasticity in the film which is introduced in the next section. The evolution of the flow resistance (yield stress) of the matrix, $\tau_{\mathrm{y}}^{\mathrm{m}}$, from the initial value, $\tau_{0}^{\mathrm{m}}$, to the saturation value, $\tau_\infty^{\mathrm{m}}$, follows the phenomenological isotropic hardening law \citep{BROWN198995, bronkhorst1992polycrystalline},
%%-----------------------------------------------------
\begin{align}\label{HarEvolMat}
\dot{\tau}_{\mathrm{y}}^{\mathrm{m}}= \dot{\gamma}^{\mathrm{m}} \, h_0^{\mathrm{m}} \, \left| 1- \dfrac{\tau_{\mathrm{y}}^{\mathrm{m}}}{\tau_\infty^{\mathrm{m}}}\right|^{a}\, \mathrm{sign}\left( 1- \dfrac{\tau^{\mathrm{m}}_{\mathrm{y}}}{\tau_\infty^{\mathrm{m}}}\right),
\end{align}
%%-----------------------------------------------------
	with $h_0^{\mathrm{m}}$ a modulus which characterize the initial hardening, and $a$ being the hardening shape factor. The evolution of the hardening in the matrix, given in Eq. \ref{HarEvolMat}, indicates that only self hardening of the matrix is accounted for here. The latent hardening between matrix and film could in principle be included here in cases where such interactions are expected to be relevant.

%------------------------------------------------------------------------------------------------------------------------------------------------------------------------
\subsubsection{Visco-plasticity of the film}
\label{sub:filmplasticity}
	Given the fact that the films are considered to be infinitesimally thin, e.g. $5-15nm$, relative to lath martensite, $50-400nm$, we formulate their response as a true sliding on discrete planes. Lets $\mathbf{n}_0$ be the normal to the embedded thin films in the reference configuration. The shear traction $\tau^{\mathrm{f}}$ on this plane can be computed from the full stress tensor, $\mathbf{M}$, applied on the system (in the lattice preserving intermediate configuration) by projecting it onto the plane as follows, 
%%-----------------------------------------------------
\begin{align}\label{Eq:6}
\tau^{\mathrm{f}}= |\mathbf{n}_{0} \cdot \mathbf{M}  \cdot(\mathbf{I} - \mathbf{n}_0 \otimes \mathbf{n}_0 ) |,
\end{align}
%%-----------------------------------------------------
	in which $\mathbf{I}$ is the $2^{nd}$ order identity tensor and $|*|$ is the \textsc{Euclidean} norm. The sliding along the plane is assumed to be in the direction of the shear traction vector, $\tau^{\mathrm{f}}$, i.e. in the direction given by,
%%-----------------------------------------------------
\begin{align}\label{Eq:61}
\mathbf{s}_{0}= \dfrac{\mathbf{n}_{0} \cdot \mathbf{M}  \cdot(\mathbf{I} - \mathbf{n}_0 \otimes \mathbf{n}_0 )}{\tau^{\mathrm{f}}}
\end{align}
%%-----------------------------------------------------
The projected shear traction, $\tau^{\mathrm{f}}$, drives the relative displacement, $v$, on the planar part of the model via a power-law relation similar to Eq. \ref{ShearMatrix},
%%-----------------------------------------------------
\begin{align}\label{ShearFilm}
\dot{v}=\dot{v}_0\left(\dfrac{\tau^{\mathrm{f}}}{\tau_{\mathrm{y}}^{\mathrm{f}}}   \right)^{1/n} ,
\end{align}
%%-----------------------------------------------------
	where $\tau_{\mathrm{y}}^{\mathrm{f}}$ is the flow resistance of the planar mode, and $\dot{v}_0$ is the reference sliding velocity. Note that for simplicity we have assumed the same rate sensitivity $n$ as for the matrix. The gliding resistance evolution on the plane reads:
%%-----------------------------------------------------
\begin{align}\label{HarEvolFil}
\dot{\tau}_{\mathrm{y}}^{\mathrm{f}}= \dot{v} \,  k_0 \, \left| 1- \dfrac{\tau_{\mathrm{y}}^{\mathrm{f}}}{\tau^{\mathrm{f}}_\infty}\right|^{a}\, \mathrm{sign}\left( 1- \dfrac{\tau_{\mathrm{y}}^{\mathrm{f}}}{\tau^{\mathrm{f}}_\infty}\right).
\end{align}
%%-----------------------------------------------------
	The flow resistance of the planar system evolves from the initial value $\tau_{0}^{\mathrm{f}}$ to the saturation value $\tau_\infty^{\mathrm{f}}$. The latent hardening due to activity of the matrix is not considered. Finally, the plastic velocity gradient due to sliding of the periodic family of films is given by,
%%-----------------------------------------------------
\begin{equation}\label{Eq:LpFilm}
\mathbf{L}_{\mathrm{P}}^{\mathrm{f}}= \dfrac{\dot{v}}{l}\mathbf{s}_0\otimes \mathbf{n}_0.
\end{equation}
%%-----------------------------------------------------
with $l$ being the spacing of the films.

%=======================================================================================================================================================================
\section{Characterization of the anisotropic effective plastic response and validation against DNS} 
\label{sec:ModelBehavior}
%------------------------------------------------------------------------------------------------------------------------------------------------------------------------
\subsection{Reference model}
\begin{figure}[ht!]
\centering
  \includegraphics[width=0.8\linewidth]{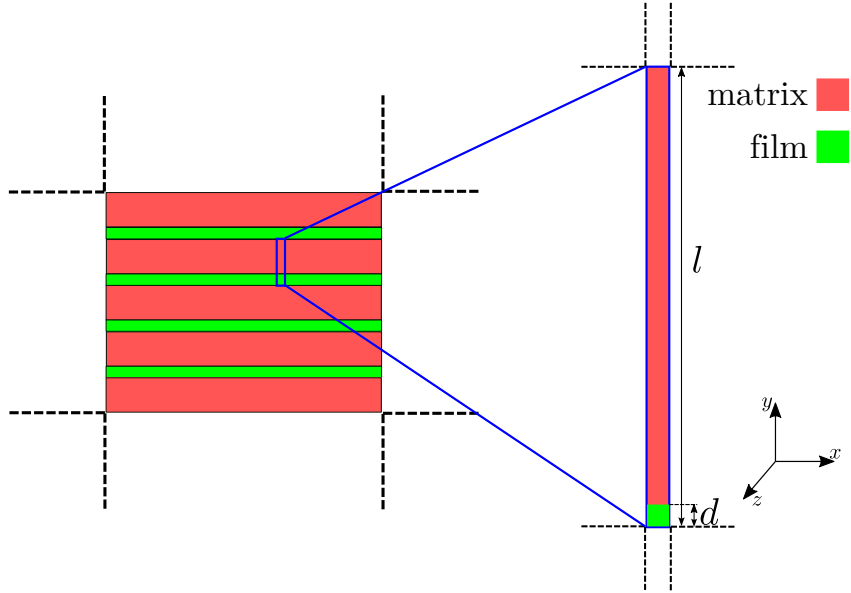}
  \caption{The periodic cell representing the reference model -- an infinite laminate with alternating layers representing the matrix and embedded thin films. The dashed lines indicate the periodicity imposed on the geometry. The volume fraction of the film is $\varphi=d/l= 0.05$. The reference model is used to assess the response of the proposed laminate model.}
  \label{fig:RVE_DNS}
\end{figure}

	The formulation of the model given in Section \ref{sec:ModelFormulation} leads to an overall anisotropic response, and the degree of anisotropy depends on the degree of activation of the planar sliding mode. Here, we aim at characterizing the inherited anisotropy and assess the impact of the assumptions made. For this purpose, a reference model is defined which represents an arbitrary infinite laminate with alternating two phases, as shown in Figure \ref{fig:RVE_DNS}. 
%%-----------------------------------------------------
%%-----------------------------------------------------
	For the reference model, direct numerical simulations (DNS) are carried out to obtain the "actual" behavior of a matrix with embedded thin films. The constitutive law for both phases of the laminate is standard visco-isotropic plasticity, and thus, their behavior separately follows Eqs. (\ref{eq:1}) -- (\ref{HarEvolMat}), without the second term in Eq. \ref{VelocityGradient}. This results in the homogeneous deformation and stress fields within each phase. With this assumption, the volume averaged mesoscale deformation gradient, $\mathbf{\overline{F}}$, and the first \textsc{Piola-Kirchhoff} stress, $\mathbf{\overline{P}}$, are obtained by the rule of mixtures,
%%-----------------------------------------------------
\begin{align}\label{RuloMixtures}
&\mathbf{\overline{F}}=(1-\varphi)\mathbf{F}^{\mathrm{m}} + \varphi\mathbf{F}^{\mathrm{f}},\\
\label{RuloMixtures2}
&\mathbf{\overline{P}}=(1-\varphi)\mathbf{P}^{\mathrm{m}} + \varphi\mathbf{P}^{\mathrm{f}},
\end{align}
%%-----------------------------------------------------
	where $\varphi$ represents the volume fraction of the soft films, which we consider to be small. The local-global interaction relations given in Eqs.\ref{RuloMixtures} $\&$ \ref{RuloMixtures2} are imposed to relate the average mechanical behavior of each composite phase to the macroscopically applied boundary conditions. This ensures satisfaction of the \textsc{Hill-Mandel} condition, stating the equivalence between the virtual internal work on the microscopic and macroscopic scale, i.e. $\langle \mathbf{P} : \delta \mathbf{F} \rangle = \mathbf{\overline{P}} : \delta \mathbf{\overline{F}}$, with $\langle \rangle$ indicating the volume average over the periodic cell \citep{HILL1963357, Mandel}. Lets $\mathbf{n}_0$ be the normal to the interface between the two phases. Kinematic compatibility requires the in-plane deformation gradient to be continuous across the interface,
%%-----------------------------------------------------
\begin{align}\label{KinComp}
\mathbf{F}^{\mathrm{f}}\cdot(\mathbf{I}- \mathbf{n}_0 \otimes \mathbf{n}_0)=\mathbf{F}^{\mathrm{m}}\cdot(\mathbf{I}- \mathbf{n}_0 \otimes \mathbf{n}_0).
\end{align}
%%-----------------------------------------------------
	Furthermore, equilibrium requires traction continuity across the interface,
%%-----------------------------------------------------
\begin{align}\label{equilib}
\mathbf{P}^{\mathrm{f}} \cdot \mathbf{n}_0 =\mathbf{P}^{\mathrm{m}}\cdot \mathbf{n}_0.
\end{align}
%%-----------------------------------------------------
	In the DNS, the film has a finite thickness, whereas the effective model of Section \ref{sec:ModelFormulation} assumes that the films are infinitesimally thin. To be able to compare the responses of the reference model and the laminate model, the film thickness $d$ in the reference model must be taken into account. Using it, the sliding velocity $\dot{v}$ in Eq. \ref{ShearFilm} may be related to the shear strain rate $\dot{\gamma}^{\mathrm{f}}$ in the reference model via, 
%%-----------------------------------------------------
\begin{align}
\dot{v} = d\dot{\gamma}^{f}= \varphi l\dot{\gamma}^{f}.
\end{align}
%%-----------------------------------------------------	
	Accordingly, the parameters $\dot{v}_{0}$ and $k_{0}$ are related to $\dot{\gamma}^{\mathrm{f}}_{0}$ and $h_{0}^{\mathrm{f}}$ via $\dot{v}_{0}=\varphi l \dot{\gamma}^{\mathrm{f}}_{0}$ and $k_{0}= h_{0}^{\mathrm{f}}/(\varphi l)$. The volume fraction of the films is taken to be $\varphi=d/l=0.05$, see Figure \ref{fig:RVE_DNS}. The material parameters of the two phases in the model are given in Table \ref{tab:PhaseBeh}. The mechanical phase contrast between the matrix and the planar mode is $\dfrac{\tau_{0}^{\mathrm{m}}}{\tau_{0}^{\mathrm{f}}}=\dfrac{\tau_{\infty}^{\mathrm{m}}}{\tau_{\infty}^{\mathrm{f}}}=2$. For simplicity, ideal plasticity is modelled by taking $h_0=0$.

%------------------------------------------------------------------------------------------------------------------------------------------------------------------------
\subsection{Comparison of shear and tensile responses}\label{subsec:ShearvsTenile}    
	The responses of the models are now investigated under two applied load cases; i) simple shear parallel to the films, and ii) uniaxial tension applied perpendicular to the films. The tension applied parallel or perpendicular to the films is expected to result in the same overall response for the reduced laminate model.
%%-----------------------------------------------------
\begin{table}[ht!]
  \caption{Model parameters used for simulating the response of the reference and the proposed model. The elastic moduli of the film are relevant only for the reference model.}
  \centering
  \begin{threeparttable}
  \begin{tabular}{cc@{\qquad}cc}
                     Parameter & film & matrix &  \\ \midrule\midrule     
        \makecell{$E$} & $210$ & $210$ & $GPa$ \\      
        \makecell{$\nu$} & $ 0.3 $ & $0.3$ & $-$ \\   \cmidrule(l r){1-4}
        \makecell{$\dot{\gamma}_{0}$} & $0.001$ & $0.001$ & $1/s$\\      
        \makecell{$\tau_{0}$} & $200$ & $400$ & MPa \\
        \makecell{$\tau_{\infty}$} & $3 \tau_{0}^{f}$ & $3 \tau_{0}^{m}$& MPa \\ 
        \makecell{$h_{0}$} & $0$ & $0$ & MPa \\   
        \makecell{$m$} & $0.02$ & $0.02$ & -- \\      
        \makecell{$a$} & $1.5$ & $1.5$ & -- \\      \midrule\midrule
  \end{tabular}
  \end{threeparttable}
  \label{tab:PhaseBeh}
  \end{table}
%%-----------------------------------------------------	
\begin{figure}[ht!]
\centering
  \includegraphics[width=0.95\linewidth]{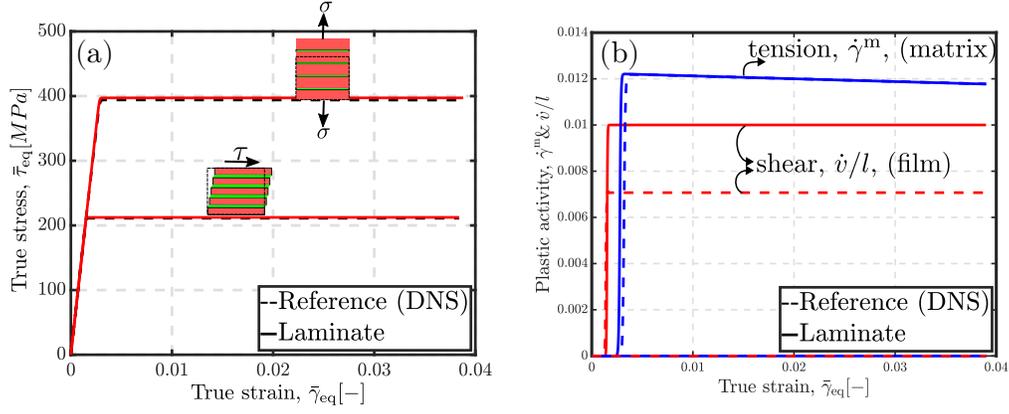}
  \caption{a) The macroscopic stress-strain response of the laminate model, plotted by solid lines, compared to that of the reference model, plotted by dashed lines. b) Plastic activity of the two modes in the laminate model, i.e. matrix and the film. Both models are subjected to simple shear parallel to the films and uniaxial tension perpendicular to the films. The volume fraction of the films in the DNS is $\varphi=0.05$. }
  \label{fig:LaminateVsDNS_Sts_Stn}
\end{figure} 
%%-----------------------------------------------------

    Figure \ref{fig:LaminateVsDNS_Sts_Stn}a compares the stress-strain behavior of the reference model, plotted by dashed lines, and the effective laminate model, plotted by solid lines, for two loading conditions. For both load cases, the response of the laminate model follows the reference model with an error of less than $2 \%$ for the predicted stress value. It is shown that both models are softer by a factor of 2 in simple shear compared to the tension case. As the shear is applied parallel to the soft films, all the plasticity in the system is carried by the films (considering ideal plasticity) and the matrix remains elastic. Therefore, the system yields at approximately the yield strength of the films, i.e. $\tau_0^{\mathrm{f}}= 200$ MPa and stays at the same stress level due to the absence of hardening. The computed yield stress is slightly ($10\%$) higher than the input value of $200$ MPa due to the rate dependence of the models. Note that the rate of deformation experienced by the films is $1/\varphi$ times higher than what is experienced by the matrix. In tension perpendicular to the films (and also parallel to the films - not shown), on the contrary, no shear stress acts on the films, which are therefore not activated. Hence, the properties of the matrix determine the behavior of the system in this direction, and an equivalent yield stress of approximately $\tau_{\mathrm{y}}^{\mathrm{m}}= 400$ MPa is observed. This explanation is supported by Figure \ref{fig:LaminateVsDNS_Sts_Stn}b, which shows the plastic activity of the two plastic modes in the laminate model, yielding of the matrix and film, throughout the applied deformation. It is shown that the plastic activity of the matrix is zero in simple shear, whereas in tension perpendicular to the films, it is the planar glide mechanism that is totally inactive.

%%---------------------------------------------------------------------------------------------------------------------------------------------------------------------- 
\section{Orientation-dependent yielding}
\label{subsec:OriDepYiel}  
\subsection{Influence of the film orientation}   
    Next, we investigate in more detail the influence of the film orientation with respect to the applied deformation. As the yield point of the effective laminate model, we define the instant at which,
%%-----------------------------------------------------
\begin{align}\label{Eq:YCriteria}
\dot{\gamma}^{\mathrm{m}}+\dfrac{d}{l}\dot{v}  \geqslant 0.002. 
\end{align}
%%-----------------------------------------------------
	The advantage of this criterion is that it is insensitive to the choice of material parameters in the model, such as the reference shear rate or hardening parameters. Accordingly, the following yield criterion is adopted for the reference model:
%%-----------------------------------------------------
\begin{align}\label{Eq:YCriteria2}
\dot{\gamma}^{\mathrm{m}}+\varphi\dot{\gamma}^{\mathrm{f}} \geqslant 0.002. 
\end{align}
%%-----------------------------------------------------
	  
	To explore the anisotropy of the effective yield surface, we apply tensile loading in the x-direction and vary the orientation of the films to cover the entire range of angles, $\theta$, between the film normal, $\mathbf{n}_{0}$, and tensile direction. The applied tensile loading in x-direction is defined by:
%%-----------------------------------------------------
\begin{equation}\label{Ch4--TensionLoad}
\mathbf{\dot{\overline{F}}}=\begin{bmatrix}
\dot{\lambda}  & \ast  & \ast \\
0     & \ast  & \ast \\
0     & 0     & \ast
\end{bmatrix}, \, \mathbf{\overline{P}}=\begin{bmatrix}
\ast  & 0     & 0   \\
\ast  & 0     & 0 \\
\ast  & \ast  & 0
\end{bmatrix}, 
\end{equation}
%%-----------------------------------------------------
	where $\overline{\square}$ denotes that the quantity below is applied on average, $\dot{\lambda}$ is the stretch rate, and $\ast$ indicates that the particular component of the tensor is free to evolve. 
	
\begin{figure}[ht!]
\centering
  \includegraphics[width=0.75\linewidth]{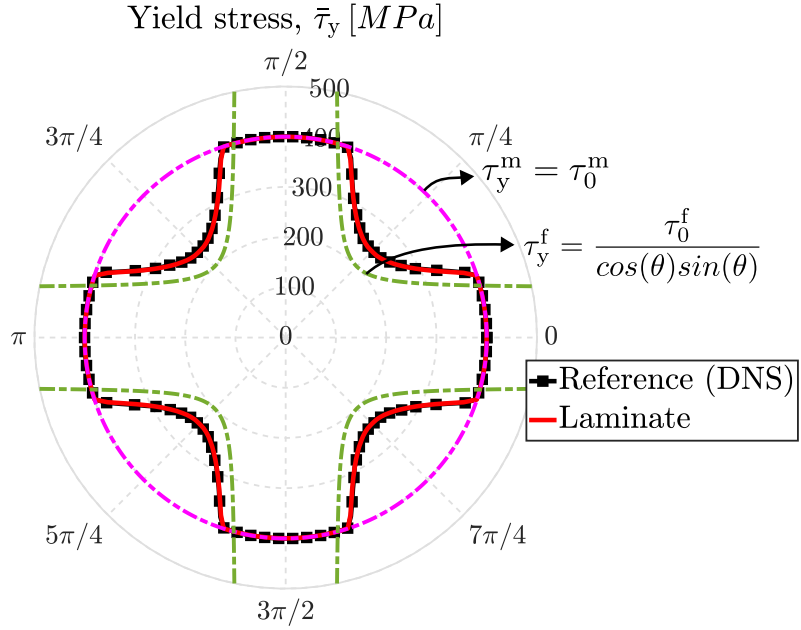}
  \caption{Polar plot comparing the initial yield stress of the effective laminate model, shown by red solid line, and the reference model, shown by black solid line overlaid by squares ($\blacksquare$). The purple dashed circle indicate the stress at which the matrix phase starts to yield, i.e. $\tau_{\mathrm{y}}^{\mathrm{m}}=\tau_{\mathrm{0}}^{\mathrm{m}}$, whereas the green dashed lines indicate the initial orientation-dependent yield strength of the films computed as, $\tau^{\mathrm{f}}_{\mathrm{y}} = \dfrac{\tau_{0}^{\mathrm{f}}}{\mathrm{cos}(\theta)\mathrm{sin}(\theta)}$. The radial axis is the yield stress in MPa, and the angular axis indicates the angle between applied tension and the normal to the films. The volume fraction of the films in the DNS is $\varphi=0.05$.}
  \label{fig:DNSvsLaminate}
\end{figure}

    Figure \ref{fig:DNSvsLaminate} shows the polar plot of the yield limit obtained by the loading condition and the yield criteria discussed above. In the plot, the radius indicates the yield stress in MPa, and the angular coordinate indicates the angle between the normal to the films and the applied tensile loading. For comparison purposes, the initial yield strength of the two phases in the laminate model are shown separately by the dashed lines. The stress at which the matrix phase starts to yield, $\tau_{\mathrm{y}}^{\mathrm{m}}=\tau_{\mathrm{0}}^{\mathrm{m}}=400$ MPa, is plotted by a purple dashed circle, whereas the initial yield stress of the films, that depends on the angle between the film normal and the applied tensile load, is computed via $\tau^{\mathrm{f}}_{\mathrm{y}} = \dfrac{\tau_{0}^{\mathrm{f}}}{\mathrm{cos}(\theta)\mathrm{sin}(\theta)}$, and plotted by the green dashed lines. The yield surface of the effective laminate model is shown by a red line and that of the reference model by a black solid line overlaid with square ($\blacksquare$) markers. Due to symmetry, only the top right quarter of the diagram is discussed in the following. It is shown that the yield surface of the effective laminate model follows that of the reference model quite accurately. It is further observed that the both models show an extremely anisotropic response. The yield strength of the models is maximum, $\bar{\tau}_{y}\approx \,400$ MPa, at $\theta=0$, and $\theta=\pi /2$. This is, as reported above, due to the fact that tension along or perpendicular to the films does not induce any shear on the soft films. Therefore, the films are inactive and yielding of the model is governed solely by the isotropic matrix, see the purple dashed circle in Figure \ref{fig:DNSvsLaminate}. For angles $\theta$ in the range of approximately $\theta < \dfrac{\pi}{8}$, as well as $\dfrac{3\pi}{8} <\theta$, the resolved shear stress on the films is too small to be activated significantly. The yield surface in these regions results from the isotropic matrix. However, as the angle between the film normal and the applied load reaches approximately $\theta=\pi/8$ (or $\theta=3\pi/8$), a transition occurs to a regime in which the softer films govern the onset of plasticity. As $\theta  \longrightarrow  \pi/4$ the applied load becomes more and more effective in activating the film sliding mechanism, resulting in a significant drop of the effective yield stress, see the dashed green lines in Figure \ref{fig:DNSvsLaminate}. The lowest level is observed for $\theta=\pi/4$, where $\bar{\tau}_{\mathrm{y}}\approx \tau_{\mathrm{y}}^{\mathrm{f}}=200$ MPa -- a factor of two lower than for $\theta=0$ or $\theta=\pi/2$. Note that the small deviation with respect to the yield strength of the films is due to the rate-dependence effect in the employed visco-plastic model. 
%%-----------------------------------------------------	
%%-----------------------------------------------------

\begin{figure}[ht!]
\centering
  \includegraphics[width=0.7\linewidth]{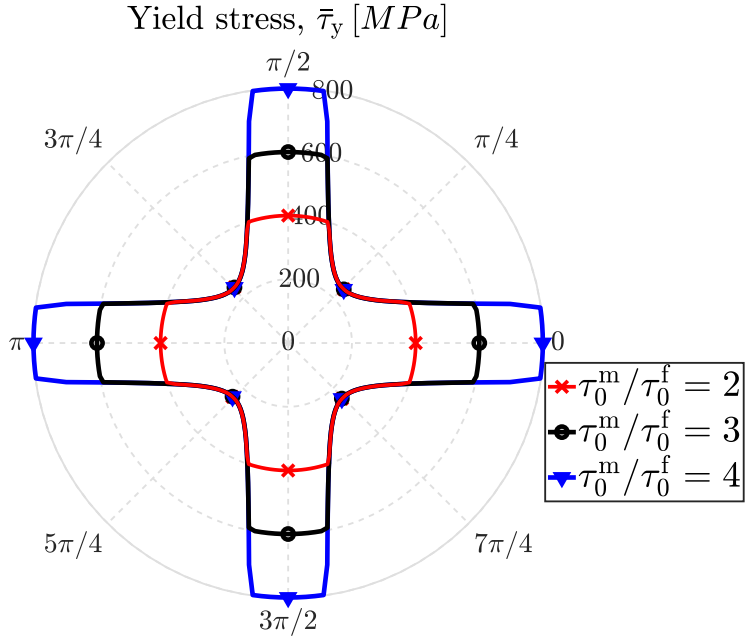}
  \caption{Polar plot comparing the yield stress of the effective laminate model for three values of the strength ratio, $\tau_{0}^{\mathrm{m}}/\tau_{0}^{\mathrm{f}}$. The volume fraction of the films is $\varphi=d/l=0.05$ .}
  \label{fig:ContrastCompare}
\end{figure}

	To assess the dependence of the yielding anisotropy on the mechanical properties of the two plastic mechanisms, the strength ratio of, $\tau_{0}^{\mathrm{m}}/\tau_{0}^{\mathrm{f}}$, is varied. This is done by increasing the initial yield strength of the matrix to $\tau_{0}^{\mathrm{m}}=600$ MPa, and $\tau_{0}^{\mathrm{m}}=800$ MPa, while keeping the initial yield strength of the films constant. Figure \ref{fig:ContrastCompare} compares the effective yield surfaces predicted by the model for these different levels of contrast.  In shear dominated load cases, i.e. around $\theta=\pi/4$, the initial yield of the model is independent of the strength of the matrix, whereas, in the region where the matrix is dominant, the yield strength of the laminate scales with the strength of the matrix. The contrast ratio $\tau_{0}^{\mathrm{m}}/\tau_{0}^{\mathrm{f}}$ hence directly controls the degree of anisotropy of the yield surface. 
%%-----------------------------------------------------	
%%-----------------------------------------------------
%%----------------------------------------------------------------------------------------------------------------------------------------------------------------------
\subsection{Influence of the film spacing and hardening}
	Next, the influence of the film spacing, $l$, on the response of the laminate is investigated. A simple shear case parallel to the films, similar to Section \ref{subsec:ShearvsTenile}, is applied, and the stress-strain response is computed for the two ratios of spacing (or volume fractions) $\varphi=d/l=0.01$ and $\varphi=d/l=0.05$. For each case of film spacing, two sets of simulations have been done: (i) with no hardening based on the parameters given in Table \ref{tab:PhaseBeh}, and (ii)  with hardening incorporated in the response of the matrix and films. In the latter case, the initial hardening modulus of the films, $k_0$, and the matrix, $h_0$, were set to, $k_{0}\,l=2\tau_0^{\mathrm{f}}$, and $h_{0}=2\tau_0^{\mathrm{m}}$, respectively.
%%-----------------------------------------------------
\begin{figure}[ht!]
\centering
  \includegraphics[width=0.7\linewidth]{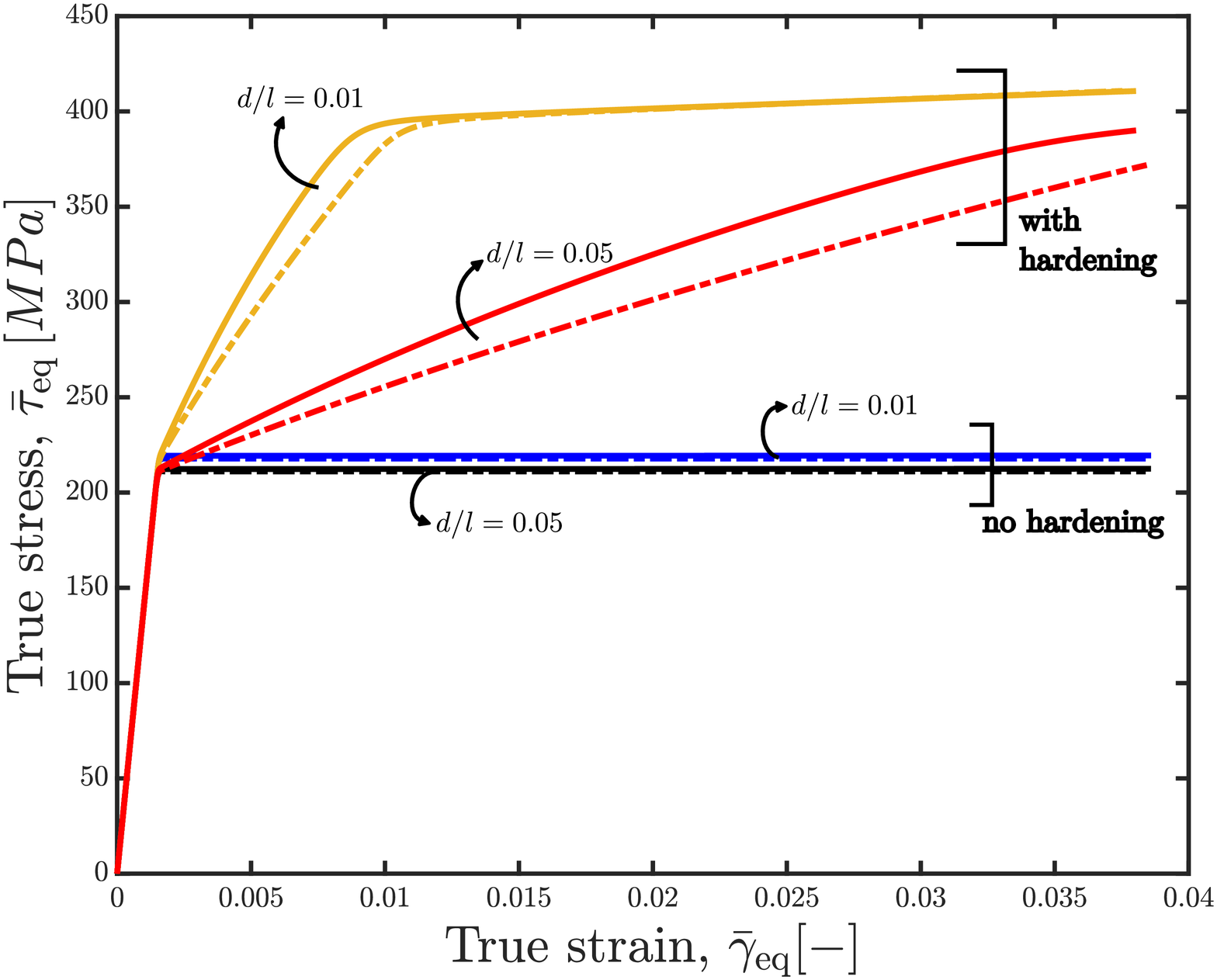}
  \caption{Macroscopic stress-strain response of the effective laminate model and the reference model under simple shear applied parallel to the films.
  The solid curves show the response of the laminate model, whereas the dashed lines show the behavior of the reference model.   
  Two different volume fractions are considered in the model, and for each case, a simulation with and without hardening was carried out. }
  \label{fig:spacing_sts_stn}
\end{figure} 
%%-----------------------------------------------------
	Figure \ref{fig:spacing_sts_stn} compares the macroscopic stress-strain responses of the laminate model, all plotted by solid lines, with the responses of the reference model, plotted by dashed lines. Four different responses are shown for each model, as we consider two different film spacings in the model (volume fraction in the case of DNS), and for each case, a simulation with and without hardening is carried out. The observations made from the Figure \ref{fig:spacing_sts_stn} are as follows. The initial yield point of both models does not change with the change in the volume fraction of the films. This is because in this particular loading, the planar film deformation mode is the softer mechanism in the system, and it starts to yield at the same stress irrespective of its corresponding volume fraction. In the case with no hardening, the responses of the two models coincide, regardless of the volume fraction of the films. However, if hardening is taken into account, some deviation is observed between the effective laminate model and the reference model. The deviation is more pronounced as the films get thicker. This is due to the assumptions incorporated in the model; the thin films are considered as planes, i.e. do not have volume, and there is no cross hardening between the systems. It is further observed that when the films are very thin, $d/l=1\%$, the secondary yield, at the yield stress of the matrix, $\tau_{0}^{m}$, is reached faster. By increasing the thickness of the films (volume fraction in DNS) to $d/l=5\%$, the transition to the secondary yield is delayed. This is intuitive since for the same applied amount of deformation, a smaller film volume will need to deform more; this leads to more hardening and hence the initial yield stress level of the matrix is reached faster.

%========================================================================================================================================================================
\section{Application to lath martensite}\label{sec:application}
\begin{figure}[ht!]
\centering
  \includegraphics[width=1\linewidth]{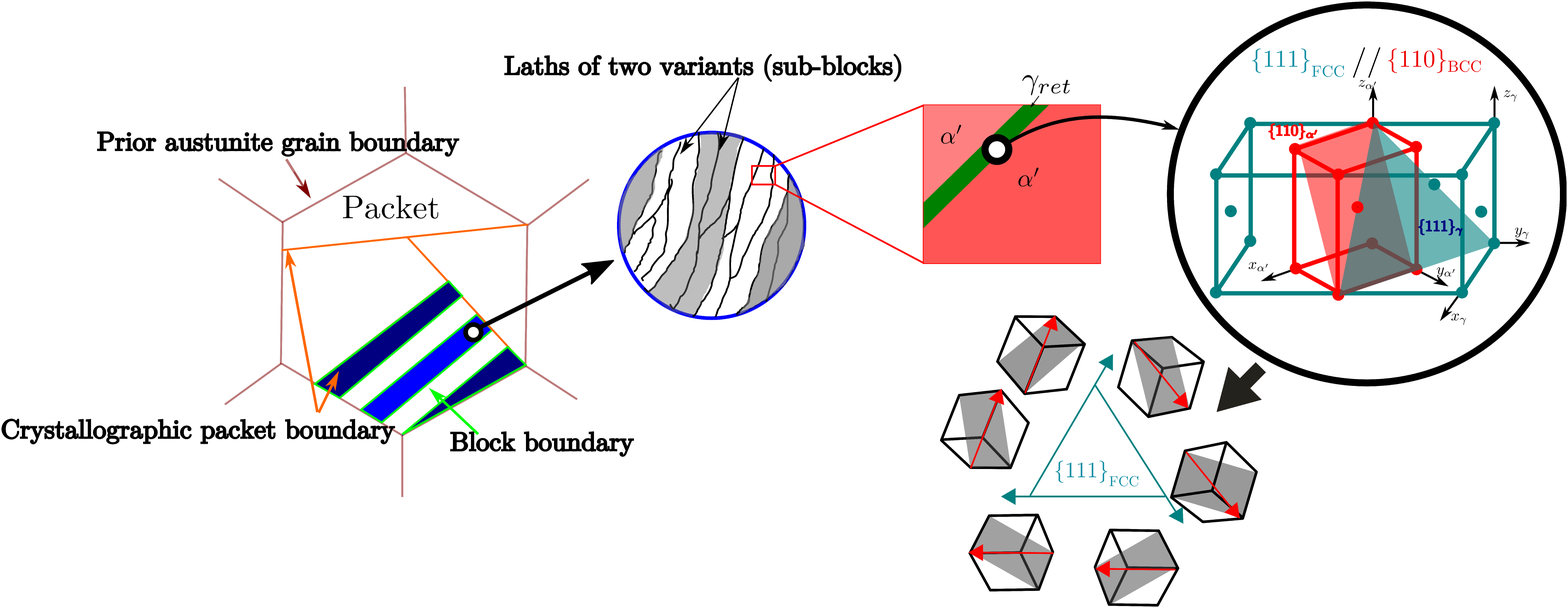}
  \caption{Schematic of an individual prior austenite grain and the lath martensite substructure formed due to the martensitic phase transformation. An illustration of six KS orientation variants is given for one packet.}
  \label{fig:SubMart}
\end{figure}

	Lath martensite is one of the main phases in advanced high strength steels, with considerable industrial significance. It has a hierarchical compound structure that forms from the austenitic phase in low-alloy steels \citep{MORITO20065323, Morito2003}. Through a diffusionless phase transformation, FCC austenite transforms to $\alpha^{\prime}$ martensite with a BCC or body-centered tetragonal (BCT) crystal structure \citep{BHADESHIA2017135}. First, at the scale of a single crystal, elongated martensite laths form. Each group of martensite laths with a particular crystallographic orientation is called a variant or sub-block. Laths of martensite with a very low misorientation constitute blocks, and several blocks sharing the same habit plane form a packet. Up to four packets of martensite can be created depending on the size of the prior austenite grain. Due to the nature of the martensitic transformation, a distinct crystallographic orientation relationship exists between the parent austenite and the transformed martensite. To describe this orientation relationship, the following expressions for parallel plane and parallel directions have been proposed, respectively: Kurdjumov-Sachs (KS) $\{111\}_{\gamma} \, || \, \{011\}_{\alpha^{\prime}}$ and $\langle 101 \rangle_{\gamma} || \langle 111 \rangle_{\alpha^\prime}$ \citep{Kurdjumow1930},  Nishiyama-Wassermann (NS) $\{111\}_{\gamma} \, || \, \{011\}_{\alpha^{\prime}}$ and $\langle 112 \rangle_{\gamma} || \langle 011 \rangle_{\alpha^\prime}$ \citep{NISHIYAMA1934, wassermann1935ueber}, and Greninger-Troiano (GT) $\{111\}_{\gamma} \, || \, \{101\}_{\alpha^{\prime}}$ and $\langle 51217 \rangle_{\gamma} || \langle 71717 \rangle_{\alpha^\prime}$ \citep{Greninger1949}. Experimental observations reveal an orientation relationship close to KS \citep{Kurdjumow1930}, in which a prior austenite grain can form $24$ different variants. The substructure of martensite and the concept of a common habit plane based on the KS orientation relationship are sketched in Figure \ref{fig:SubMart}.  
%%-----------------------------------------------------
%%-----------------------------------------------------

	 There have been experimental reports on the existence of thin austenite films retained between the laths of martensite, due to an incomplete martensitic transformation, in martensitic steels \citep{morito2011carbon, COTAARAUJO2021106445} and dual-phase (DP) steels \citep{liao2010microstructures, yoshida2015crystallographic}. These retained austenite films are expected to be softer than the martensite laths. It has been argued that they may facilitate an inter-lath boundary sliding mechanism \citep{Maresca2014,maresca2016reduced}, akin to the film sliding in the effective laminate model presented here. Accordingly, we apply the model in this section to lath martensite with retained austenite films.

%-----------------------------------------------------------------------------------------------------------------------------------------------------------------------
\subsection{Reference model}\label{sec:RefModelCP}
	   
	A reference computational model has been set up that consist of a periodic two-phase laminate composed of an alternating stack of a BCC single crystal (lath martensite) as the matrix phase and an FCC single crystal (retained austenite) as the embedded thin film. For the reference model, the same periodic cell shown in Figure \ref{fig:RVE_DNS} is used to account for the two adjacent crystals present in the system. The ratio of the thickness of the austenite films to the thickness of the martensite laths is reported in the literature to be $4\% - 10\%$ \citep{morito2011carbon, liao2010microstructures}. Here, consistent with the previous section, we take the volume fraction of austenite to be $\varphi = 0.05$. The KS orientation relationship for variant $1$, $\{111\}_{\gamma} \, || \, \{011\}_{\alpha^{\prime}}$ parallel plane, and $\langle \bar{1}01 \rangle_{\gamma} || \langle \bar{1}\bar{1}1 \rangle_{\alpha^\prime}$ parallel direction is taken into account \citep{Morito2003}. This is done by aligning the direction of the variant parallel to the x-direction and perpendicular to the y-direction. 
	
	The plasticity of the crystal is assumed to be described by a standard crystal plasticity formulation given by \citep{peirce1983material, bronkhorst1992polycrystalline}: 
%%-----------------------------------------------------
\begin{align}\label{LPCP}
\mathbf{L}_{\mathrm{p}}=\sum_{\alpha=1}^{n_s} \dot{\gamma}^{\alpha} \mathbf{P}_{0}^{\alpha},
\end{align}
%%-----------------------------------------------------
where $\mathbf{P}_{0}^{\alpha}= \mathbf{s}_0^{\alpha}\otimes \mathbf{n}_0^{\alpha}$ is the \textsc{Schmid} tensor related to the $\alpha$th slip system, obtained by the dyadic product of $\mathbf{s}_0^{\alpha}$, the slip direction, and $\mathbf{n}_0^{\alpha}$, the normal to the slip plane in the intermediate configuration. Similar to Eq. \ref{ShearFilm}, the plastic slip rate on the active slip systems, $\dot{\gamma}^{\alpha}$, is given by a rate-dependent power-law relation,
%%-----------------------------------------------------
\begin{align}\label{ShearCP}
\dot{\gamma}^{\alpha}=\dot{\gamma}_0^{\alpha}\left(\dfrac{|\tau^{\alpha}|}{s^{\alpha}}   \right)^{1/m}\mathrm{sign}(\tau^{\alpha}),
\end{align}
%%-----------------------------------------------------
in which $\dot{\gamma}_0^{\alpha}$ denotes the reference slip rate, and $m$ is the rate sensitivity factor. The plastic deformation is governed by the resolved shear stress, $\tau^{\alpha}=(\mathbf{S}^{\mathrm{dev}}_\mathrm{e} \cdot \mathbf{C}_\mathrm{e}): \mathbf{P}_{0}^{\alpha}$, on the $\alpha^{\mathrm{th}}$ slip system. In order to incorporate hardening in the model, an evolution equation for the slip resistance, $s^{\alpha}$, is formulated as, 
%%-----------------------------------------------------
\begin{align}\label{Eq:EvolS}
\dot{s}^{\alpha}= \sum_{\beta=1}^{N_s}h^{\alpha\beta}|\dot{\gamma}^{\beta}|.
\end{align}
%%-----------------------------------------------------
Note that $s^{\alpha}$ evolves from an initial resistance, $s_{0}$, to a saturation value, $s_{\infty}$, with the hardening modulus, $h^{\alpha\beta}$, evolving due to self hardening of the slip system $\alpha$ and latent hardening induced by other systems, $\beta$, given as,
%%-----------------------------------------------------
\begin{align}\label{Eq:HardenS}
h^{\alpha\beta} = h_{0} \left( 1-\dfrac{s^{\alpha}}{s_{\infty}} \right)^{a} (q+(1-q)\delta^{\alpha\beta}).
\end{align}
%%-----------------------------------------------------
	Here, $h_{0} $ denotes the reference hardening modulus, $q$ is the latent hardening ratio, and $\delta^{\alpha\beta}$ is the Kronecker delta. For the martensite lath, two cases with either one $\{110\}_{\alpha^\prime}$ or two $\{110\}_{\alpha^\prime}\, \& \, \{112\}_{\alpha^\prime}$ slip families are simulated. The same parameter set is used for both BCC slip families. The material parameters of the BCC and FCC crystals are given in Table \ref{tab:CPBeh}. 
%%-----------------------------------------------------
\begin{table}[ht!]
  \caption{Material parameters used in the simulations for bicrystal of lath martensite and retained austenite.}
  \centering
  \begin{threeparttable}
\begin{tabular}{cc@{\qquad}cc}
      property & FCC & BCC & unit \\ \midrule\midrule     
        \makecell{$E$} & $210$  & $210$ & $GPa$ \\      
        \makecell{$G$} & $85$ & $85$ & $GPa$ \\     \cmidrule(l r){1-4}
        \makecell{$\dot{\gamma}_{0}$} & $0.001$ & $0.001$ & $s^{-1}$\\      
        \makecell{$s_{0}$} & $200$ & $400$ & MPa \\     
        \makecell{$m$} & $0.02$ & $0.02$ & -- \\      
        \makecell{$a$} & $1.5$ & $1.5$ & -- \\      \midrule\midrule
    \end{tabular}
  \end{threeparttable}
  \label{tab:CPBeh}
  \end{table}
%%-----------------------------------------------------

%-----------------------------------------------------------------------------------------------------------------------------------------------------------------------
\subsubsection{Orientation factor}
	The constitutive parameters used in the effective laminate model are similar to the ones given in Table \ref{tab:PhaseBeh}. However, to link the microscale BCC slip system properties to the homogeneous isotropic properties of the matrix, the well-known \textsc{Taylor} orientation factor $T$ has to be introduced in the model \citep{rosenberg1971calculation}. For BCC crystals, this factor is reported to be $T^{\mathrm{m}}=2.45$, which is an average value for a polycrystal of many grains obtained based on the CPFEM estimation with one slip family subjected to uniaxial tensile load \citep{zhang2019assessment}. However, the formulation of the models given in this paper is based on the shear equivalent stress value, i.e. $\tau_{\mathrm{eq}}= \sqrt{\dfrac{1}{2} \mathbf{M}^{\mathrm{dev}}:\mathbf{M}^{\mathrm{dev}}}$. Accordingly, we employ an orientation factor equal to  $T^{\mathrm{m}}=2.45/\sqrt{3}$. This factor enters the isotropic visco-plasticity formulation of the matrix phase described in Section \ref{sub:matrixplasticity}, by modifying the plastic velocity gradient as follows,
%%-----------------------------------------------------
\begin{align}\label{Lp3DOri}
\mathbf{L}_{\mathrm{P}}^\mathrm{m}= \dfrac{\dot{\gamma}^{\mathrm{m}}}{T^{\mathrm{m}}}\dfrac{\mathbf{M}^{\mathrm{dev}}}{|| \mathbf{M}^{\mathrm{dev}} ||_{F}}. 
\end{align}
%%-----------------------------------------------------	
While, the plastic shearing rate is modified to,
%%-----------------------------------------------------
\begin{align}\label{gammaOri}
\dot{\gamma}^{\mathrm{m}}=\dot{\gamma}_0\left( \sqrt{\dfrac{1}{2}}  \dfrac{ || \mathbf{M}^{\mathrm{dev}} ||_{F}}{T^{\mathrm{m}}\tau_{\mathrm{y}}^{\mathrm{m}}} \right)^{1/m}.
\end{align}
%%-----------------------------------------------------	

	For similar reasons, an orientation factor must be taken into account for linking the FCC slip system properties on one slip plane to the planar isotropic plasticity model of the films. Note, however, that the situation is slightly different here because the orientation of the slip plane associated with the relevant slip systems is identical to the orientation of the films, and hence the effect of random slip plane orientation does not need to be accounted for in a \textsc{Taylor}-like factor. What remains to be accounted for is the random orientation of the slip directions in that plane. Without an orientation factor, the planar film model would assume that a slip system is always available in exactly the direction of maximum shear traction. In reality, and in the crystal plasticity model, the "best available" slip direction may be at some angle with respect to the shear traction, resulting in a slightly harder response. The orientation factor for the film, $T^{\mathrm{f}}$, accounts, on average, for this misalignment. It has been determined by computing the average response of a FCC plane with 3 slip systems oriented $60^\circ$ with respect to each other under an applied simple shear load parallel to the plane.  Consequently, the average orientation factor has been computed to be $T^{\mathrm{f}}=1.1$. This factor modifies the constitutive law describing the behavior of the thin films given in Section \ref{sub:filmplasticity} by adjusting the relative displacement to,
%%--------------------------------------------
\begin{align}\label{ShearFilmT}
\dot{v}=\dot{v}_0\left(\dfrac{\tau^{\mathrm{f}}}{T^{\mathrm{f}}\tau_{\mathrm{y}}^{\mathrm{f}}}   \right)^{1/m} ,
\end{align}
%%--------------------------------------------
and accordingly the corresponding velocity gradient to, 
%%--------------------------------------------
\begin{equation}\label{Eq:LpFilm2}
\mathbf{L}_{\mathrm{P}}^{\mathrm{f}}= \dfrac{\dot{v}}{T^{\mathrm{f}}l}\mathbf{s}_0\otimes \mathbf{n}_0.
\end{equation}
%%--------------------------------------------	

%-----------------------------------------------------------------------------------------------------------------------------------------------------------------------	
\subsection{Single packet response and comparison to reference model}
\begin{figure}[ht!]
\centering
  \includegraphics[width=0.7\linewidth]{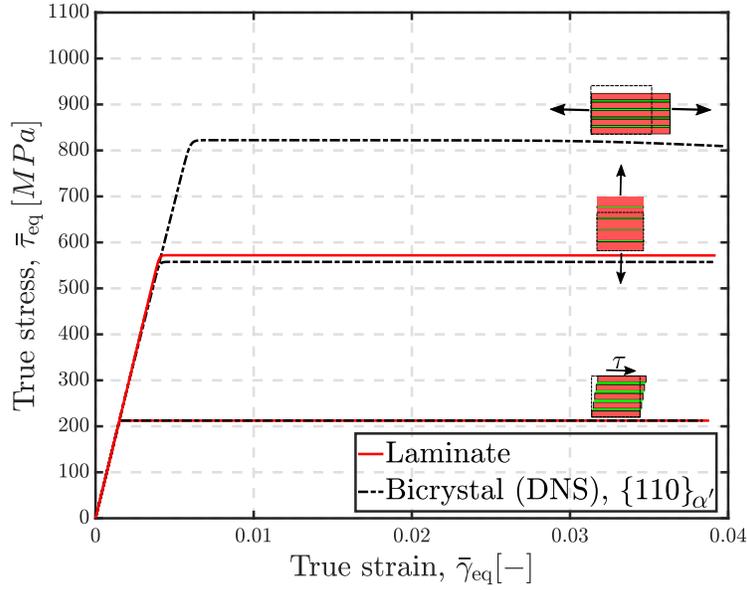}
  \caption{The stress-strain behavior of the laminate model, plotted by solid lines, is compared to the response of the reference bicrystal model, plotted by dashed lines. Three loading cases are applied; simple shear along the habit plane and tension along and perpendicular to the habit plane. The volume fraction of the films is $\varphi=0.05$ for all cases.}
  \label{fig:CPvsLaminate}
\end{figure}

	The stress-strain behavior of the laminate model, plotted by solid lines, is compared to the response of the reference bicrystal model, plotted by dashed lines, in Figure \ref{fig:CPvsLaminate}. Three different load cases are applied. First, simple shear along the habit plane in xy-direction, i.e. $\tau_{12}$, parallel to the direction of the first variant, i.e. $\langle \bar{1}01 \rangle_{\gamma} || \langle \bar{1}\bar{1}1 \rangle_{\alpha^\prime}$. Second, tension in y-direction, perpendicular to the habit plane in the direction of $\{111\}_{\gamma} \, || \, \{011\}_{\alpha^{\prime}}$. Due to the anisotropy of the bicrystal when it is pulled along and perpendicular to the $\{011\}_{\alpha^{\prime}}$ plane, the response of tension along the habit plane is plotted for the reference bicrystal model as well. The laminate model gives the same response for the two tension load cases,  i.e. the single red curve in Figure \ref{fig:CPvsLaminate} for both.
%%-----------------------------------------------------
%%-----------------------------------------------------
	Figure \ref{fig:CPvsLaminate} shows that under shear applied parallel to the interface plane of bicrystal (in the direction of the in-plane slip system in the bicrystal) the response of the both models coincide. However, under tension along and perpendicular to the habit plane, the responses are inconsistent. The main origin of the deviations is the assumption of isotropy in the matrix, requiring the \textsc{Taylor} factor in the laminate model. This factor has been calculated based on the average response of a sufficiently large polycrystal, i.e accounts for the average of all directions. In the case of the reference bicrystal model subjected to a tensile load, the FCC films are inactive, and the \textsc{Schmid} factor for each direction of the applied load for the BCC crystal is different than the incorporated averaged value, $T^{\mathrm{m}}=2.45/\sqrt{3}$, which hence leads to quantitative differences.

\begin{figure}[ht!]
\centering
  \includegraphics[width=0.95\linewidth]{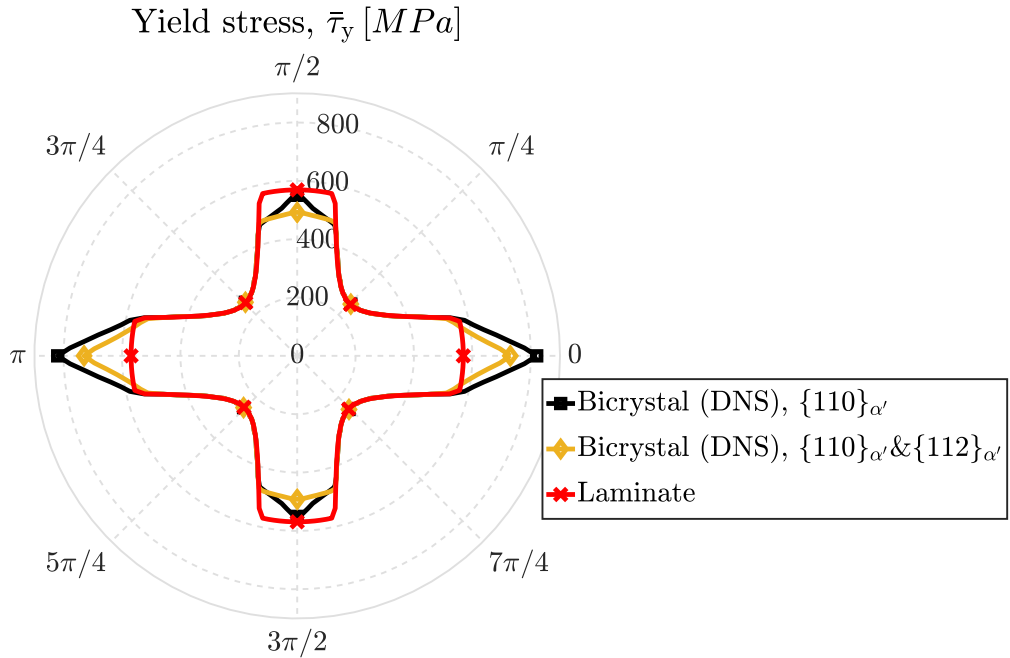}
  \caption{The polar plot compares the yield stress of the model with that of the reference bicrystal model. \textsc{Taylor} factor for the laminate model is $2.45/\sqrt{3}$. The BCC crystal is simulated once with $\{110\}$ slip family,  which is marked by black ($\square$) markers, and then with two slip families of $\{110\} \& \{112\}$, which is marked by yellow ($\diamond$) markers. The volume fraction of the films is $\varphi=5\%$ for all cases. The axes of the polar plot indicate similar quantities as in previous polar plots.}
  \label{fig:LaminateVsCP_Yield}
\end{figure}

	Figure \ref{fig:LaminateVsCP_Yield} compares the yield surface prediction of the laminate model to the prediction of the reference bicrystal model. The BCC crystal in the reference model is simulated with one and two slip system families. It is observed that in shear dominant loads, there is a close agreement between the prediction of the laminate model and the reference bicrystal model. However, in tension there remains a slight difference in the predictions, which originates from the value of the \textsc{Taylor} factor for the laminate model as explained above. 
%%-----------------------------------------------------
%%-----------------------------------------------------	

%-----------------------------------------------------------------------------------------------------------------------------------------------------------------------	
\subsection{Mesoscale simulation of lath martensite in DP steel with the effective laminate model}
\label{sec:Meso-scale}
\begin{figure}[ht!]
\centering
  \includegraphics[width=1\linewidth]{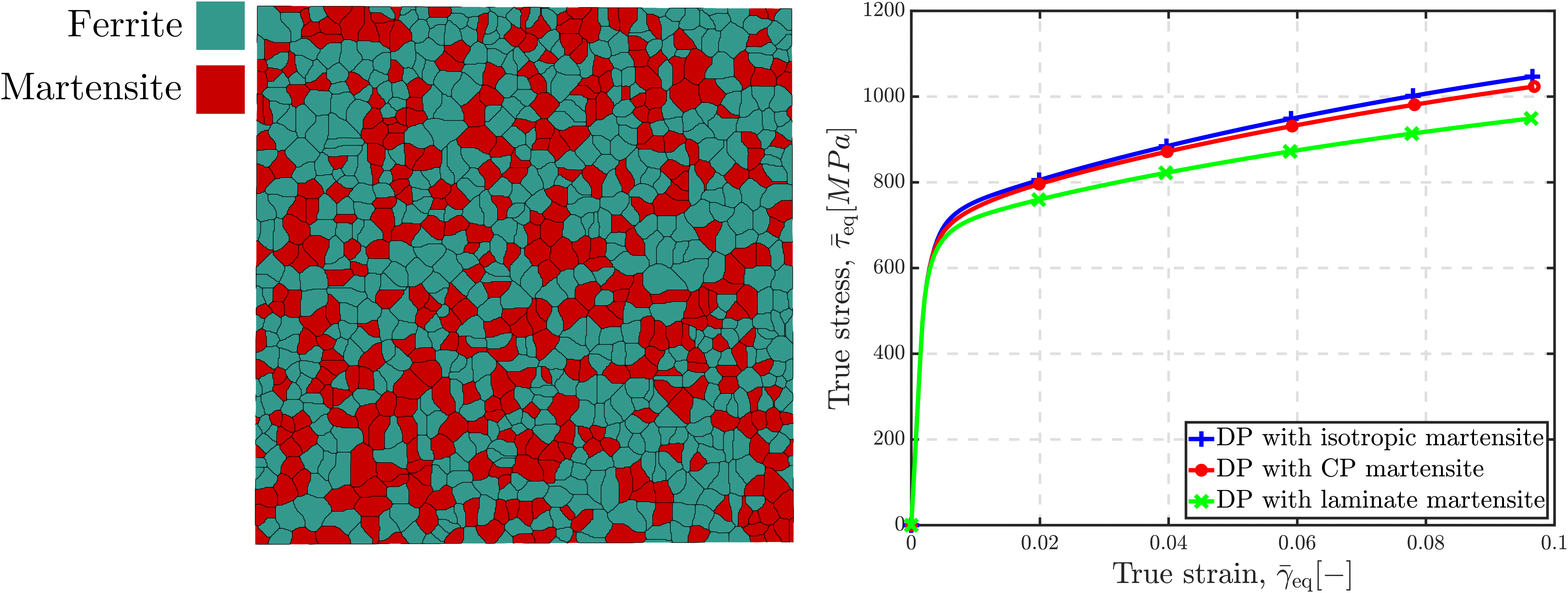}
  \caption{Left: virtually generated DP steel microstructure with martensite grains shown in red and ferrite in cyan. The grain boundaries are shown in black. The martensite volume fraction is $40\%$. A random crystallographic orientation is assigned to martensite grains, and each individual grain is considered to be a single packet, i.e. with one particular orientation for the film as used in the effective laminate model. 
  Right: the macroscopic stress-strain response of this microstructure computed with three different models incorporated in the martensite grains.}
  \label{fig:RVE40_Macro}
\end{figure}

	In the final part of this paper we apply the laminate model at the scale it is intended for: the polycrystalline mesoscale. At this scale, many individual grains and packets can be distinguished, but fully resolving their substructure is computationally prohibitively expensive. To carry out the mesoscale RVE simulations, the laminate model has been implemented in the DAMASK software \citep{ROTERS2019420}. DAMASK is a unified multi-physics spectral method-based simulation package that incorporates a variety of constitutive models and homogenization approaches. A virtual DP steel microstructure with randomly distributed ferrite and martensite grains, as shown in Figure \ref{fig:RVE40_Macro}, is considered. The volume fraction of the martensite grains is $40\%$, and each martensite grain is considered to be a single crystallographic packet, i.e., having a unique habit plane orientation. The orientation relationship between neighboring packets stemming from the same prior austenite grain is not taken into account. Three simulations are carried out wherein the martensite grains are modeled and compared using three different constitutive models: i) isotropic visco-plasticity as defined in Eqs. (\ref{eq:1}) -- (\ref{HarEvolMat}), without the second term in Eq. \ref{VelocityGradient} , ii) crystal plasticity as defined in Eqs. (\ref{LPCP}) -- (\ref{Eq:HardenS}), and iii) the effective laminate model. The crystal plasticity case is considered to investigate the effect of martensite texture on the anisotropy of the microstructure, and compare the predicted BCC-only response to that of the effective laminate model. The ferrite grains are modeled using the isotropic visco-plasticity model, as defined in Eqs. (\ref{eq:1}) -- (\ref{HarEvolMat}), without the second term in Eq. \ref{VelocityGradient}, in all of the simulations. The material parameters for ferrite, retained austenite and lath martensite are adopted from \citep{maresca2014subgrain, MARESCA201674}. The volume fraction of retained austenite is taken to be $\varphi=0.05$ which is in agreement with the reports of \citep{yoshida2015crystallographic,liao2010microstructures}, which are based on observations made for lath martensite in DP steels. A uniaxial tensile load in the x-direction, similar to Eq. \ref{Ch4--TensionLoad}, is applied to the periodic RVE in which the stretch rate, $\dot{\lambda}$, is incremented from zero to $0.1$. 
	
	Figure \ref{fig:RVE40_Macro} (right) shows the macroscopic stress-strain response computed for the three simulations done on the DP steel microstructure shown on the left. The responses of the cases in which martensite is modeled via CP and isotropic plasticity are almost identical. However, a significantly softer response is observed for the simulation with the effective laminate model. The thin soft films parallel to the habit plane of lath martensite entail softer martensite grains and hence a softer material response. Note that this mechanism is absent in both the isotropic model, and the crystal plasticity model.
%%-----------------------------------------------------
%%-----------------------------------------------------

\begin{figure}[ht!]
\centering
  \includegraphics[width=1\linewidth]{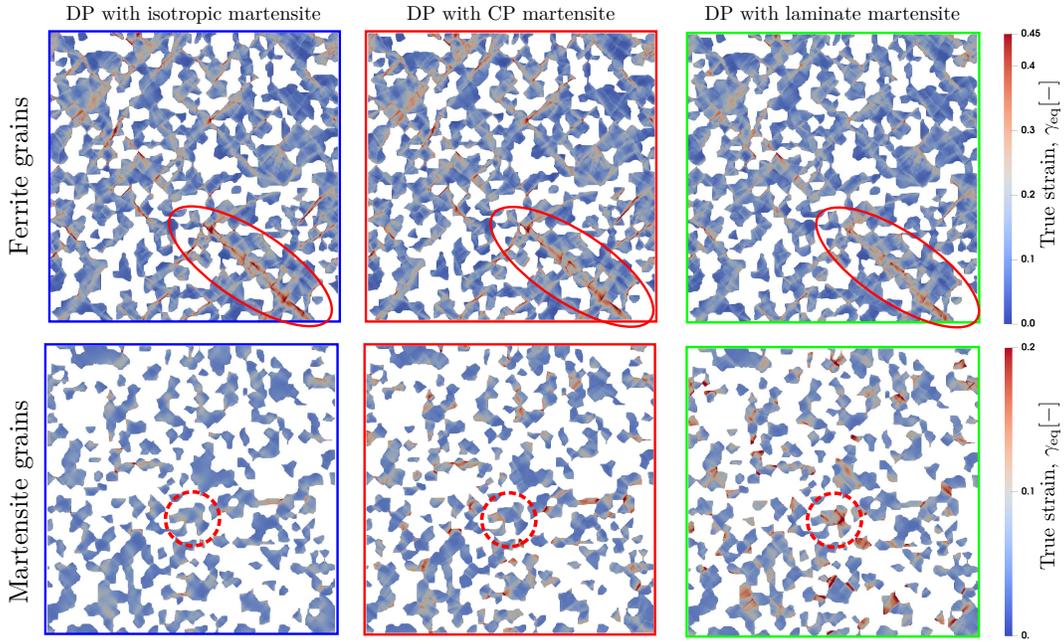}
  \caption{Computed equivalent plastic strain in the ferrite and martensite grains captured at an applied global strain of $\bar{\gamma}_{\mathrm{eq}}\approx 0.1$. The three columns show the different constitutive models used in the martensite grains. The top row shows the ferrite grains, the bottom row the martensite. In all of the simulations ferrite is modelled via the isotropic plasticity constitutive law. }
  \label{fig:DP40_Grains}
\end{figure}

	While the effect of the soft austenite films on the global response is already significant, an even more profound influence is observed on the local deformation in the microstructure. The equivalent strain map of the ferrite and martensite grains are compared for the three simulations and the results are shown in the top and bottom row of Figure \ref{fig:DP40_Grains}, respectively. The local strain maps are computed at a globally applied strain of $\bar{\gamma}_{\mathrm{eq}}\approx 0.1$. It is observed that the strain distributions in the ferrite are qualitatively similar for all the cases, i.e. the same localized areas are observed. This pattern also matches in a quantitative sense for the simulations in which the isotropic and CP models (both without retained austenite films) used in martensite. However, it is shown that by incorporating the effective laminate model in the martensite grains, the strain localization in the ferrite grains is decreased. An example of this behavior is highlighted by an ellipse around the ferrite channel in the top row of Figure \ref{fig:DP40_Grains}. As a result of activating the sliding mode in lath martensite, this phase can accommodate more plasticity in the direction of the films, leading to a decrease in the amount of localization in the ferrite grains. This observation is supported by Figure \ref{fig:DP40_Grains} bottom row, where the spatial distribution of the strain in the martensite is compared for the three cases. The highlighted region (dashed circle in bottom row) is one of the areas indicating a higher contribution of the martensite grains to plasticity when the effective laminate model is used. 
%%-----------------------------------------------------
%%-----------------------------------------------------

\begin{figure}[ht!]
\centering
  \includegraphics[width=1\linewidth]{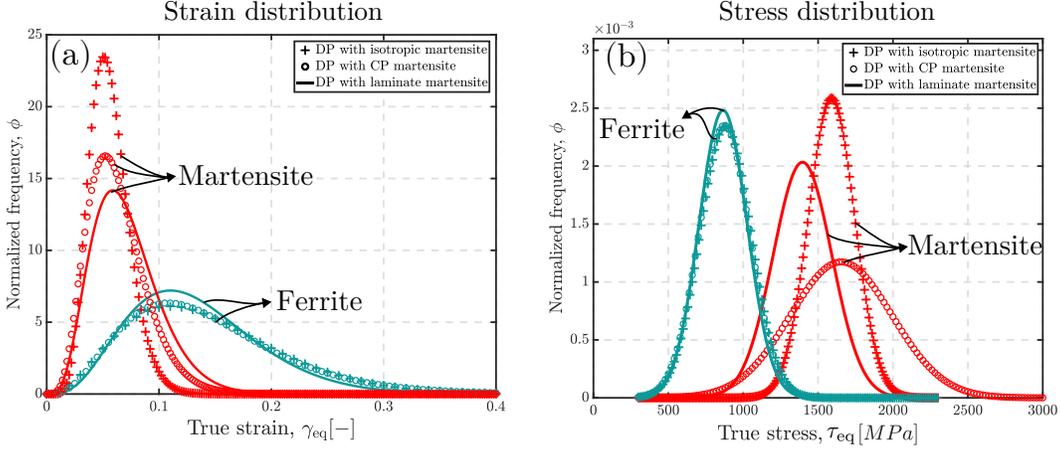}
  \caption{The probability density $\Phi$ of the a) equivalent true strain $\gamma_{\mathrm{eq}}$, and b) equivalent true stress $\tau_{\mathrm{eq}}$ MPa for the ferrite and martensite grains plotted for the three models used in lath martensite. The results are obtained at the end of the deformation, at $\bar{\gamma}_{\mathrm{eq}}\approx 0.1$.}
  \label{fig:Histogram_full}
\end{figure}

	The above observations are confirmed in Figure \ref{fig:Histogram_full}, where the local stress-strain distributions of the grains are plotted for the three simulations. It is demonstrated that when the effective laminate is used for the martensite grains, shown by solid lines, the stress-strain partitioning has reduced in the microstructure, i.e. closer mean values for the ferrite and martensite grains in both stress and strain distribution diagrams. The highest scatter of the strain distribution in the martensite grains is observed when the effective laminate is used as the constitutive choice of the martensite. This agrees with the observation made in \citep{MORSDORF2016202}-(Figure. 3), who show that the heterogeneity of the strain distribution in the martensite grains is not correlated with the heterogeneity of the corresponding crystallographic orientation distribution, i.e. the \textsc{Taylor} factor map. Moreover, Figure \ref{fig:Histogram_full} shows that the average local stress is the lowest for martensite described by the effective laminate model. This results from the embedded soft films, even though the heterogeneity of the stress distribution is higher for martensite grains modelled with CP. The significant scatter of stress for the CP case is due to variations in the \textsc{Schmid} factor; from minimum $0.22$ to maximum $0.5$ for a BCC crystal with one slip family.   
%%-----------------------------------------------------
%%-----------------------------------------------------

%-----------------------------------------------------------------------------------------------------------------------------------------------------------------------
\subsection{Computational efficiency}	
\begin{figure}[ht!]
\centering
  \includegraphics[width=0.6\linewidth]{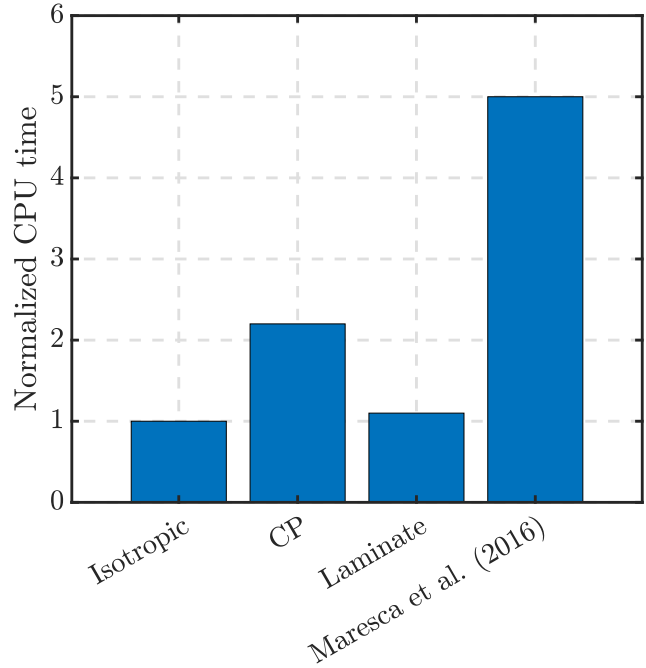}
  \caption{The normalized CPU time computed for the DP microstructure shown in Figure \ref{fig:RVE40_Macro} using different constitutive models implemented for the martensite grains. The data for CPU time of the Maresca et al. (2016) is provided by the corresponding author of \citep{MARESCA201674}, who has conducted a similar comparison. }
  \label{fig:CPUtime}
\end{figure}

	Finally, the computational efficiency of the effective laminate model is compared to the other two cases studied above. The model proposed by \citet{maresca2016reduced} is considered as well for the comparison of the CPU time, since, to our best knowledge, it is the only existing model in the literature accounting for the boundary sliding in the lath martensite which is computationally cheap enough to be applied in mesoscale simulations. In that model, the lath martensite microstructure is modelled as a lamellae composed of repeating laths of martensite and austenite films. The model incorporates isotropy for lath martensite and the out-of-habit plane systems of austenite, whereas the three in-habit plane slip systems of the retained austenite are modelled via a standard crystal plasticity approach. Figure \ref{fig:CPUtime} shows the CPU time for all of the simulations normalized by that of the isotropic plastic simulation. The corresponding author of \citep{MARESCA201674} provided the CPU time data for Maresca et al. (2016), see Figure \ref{fig:CPUtime}. The data is obtained by a similar comparison done on a DP microstructure with $33\%$ of martensite volume fraction. However, since the models are implemented in different solvers, and the efficiency of the codes can be different, the CPU time comparison is only indicative, and no sharp quantitative conclusions can be made. It is shown that the effective laminate model proposed in this paper, has a computation time comparable to that of isotropic plasticity while it preserves the physics of the boundary sliding and anisotropic plasticity observed in the martensite. The model proposed by \citet{maresca2016reduced} has a higher computational time as it addresses satisfaction of equilibrium and kinematic compatibility at each time increment between the three plastic mechanisms of the system. The effective laminate model assumes iso-stress state for the embedded films and hence equilibrium is satisfied trivially, and no local iteration check for compatibility is needed, as it is violated. In essence, since the films can only experience in-plane shear, other out-of-plane components of deformation, such normal components, will not be relevant for the films. Moreover, the anisotropy in the habit plane which is considered in the model of \citet{maresca2016reduced}, is disregarded in the effective laminate model, as it has minor influence on the strains induced by the sliding mode. 
%%-----------------------------------------------------

%======================================================================================================================================================================
\section{Conclusions}
	A computationally efficient micromechanics based model is proposed to capture the anisotropic behavior of a particular class of lamellar microstructures in which thin soft lamellae can be condensed into discrete slip planes. These thin planes are embedded in a matrix which represents the harder lamellae. Accordingly, the model can be summarized as an isotropic visco-plastic model which is enriched with an additional orientation-dependent planar plastic mode. A comparison with direct numerical simulations done on a reference model, which is a two-phase laminate sharing the same geometrical and material characteristics, shows that the model recovers the same response under shear applied parallel and tension applied perpendicular to the films. The yield surface of both models shows the highest value for tension applied parallel or perpendicular to the soft films. The weakest response is observed when the applied load entails a maximum shear stress, that aligns with the plane of the soft film. 

    The formulation is applied to model the substructure of lath martensite with inter-lath thin austenite films. It is shown that in shear-dominated cases the yield strength of the effective laminate model nicely coincides with the fully resolved model of a bicrystal consisting of lath martensite and an austenite thin film. The differences observed in tension dominated loads is due to the \textsc{Taylor} factor used to approximate an anisotropic crystal with a isotropic model incorporated in the matrix. The effective laminate model is used in mesoscale simulations of a DP microstructure, and the results are compared to the cases in which the martensite grains are modelled via isotropic plasticity and CP constitutive models, without the effect of the softer plastic mechanism aligned with the corresponding habit planes. In the case of martensite grains modelled with the effective laminate model, as a result of embedding soft films aligned with the habit plane in lath martensite, contribution of this phase to plasticity is increased, and hence, a relatively homogeneous stress-strain distribution is observed in the microstructure. For the same reason, a drop in the macroscopic yield stress of the DP microstructure is observed. The computational gain of the model is investigated, whereby it is demonstrated that the computationally efficiency of the model is almost at the same level of the isotropic visco-plastic model, while it still captures the physics of boundary sliding and extreme plastic anisotropy in the lath martensite.

\section*{Acknowledgements}
This research was carried out under project number T17019b in the framework of the Research Program of the Materials innovation institute (M2i) (\href{www.m2i.nl}{www.m2i.nl}) supported by the Dutch government.

\newpage

\bibliography{/home/ador/Desktop/Papers/3rdPaper/FinalVersion/ArXiv/main.bib}
%\bibliography{library.bib}
\bibliographystyle{unsrtnat}

\end{document}